\documentclass{aa}
\usepackage{txfonts}
\usepackage{graphicx}

\begin{document}

\title{The Resonant Dynamical Evolution of Small Body Orbits Among Giant Planets}

\author{Ryszard Gabryszewski \inst{1} and Ireneusz W{\l}odarczyk \inst{2}}
\offprints{Ryszard Gabryszewski}
\institute{Space Research Centre, ul. Bartycka 18A, 00-716 Warszawa, Poland\\ 
           \email{r.gabryszewski@cbk.waw.pl}
           \and
           Astronomical Observatory of the Chorz\'{o}w Planetarium, WPKiW, 41-501 Chorz\'{o}w, Poland\\ 
           \email{astrobit@ka.onet.pl}}

\authorrunning {R. Gabryszewski and I. W{\l}odarczyk}

\titlerunning{The Resonant Dynamical Evolution ...}

\date{Received 15 January 2003 / Accepted 25 April 2003}

\abstract{
Mean motion resonances (MMRs) can lead either to chaotic or regular motion. 
We report on a numerical experiment showing that even in one of the most 
chaotic regions of the Solar System - the region of the giant planets, there
are numerous bands where MMRs can stabilize orbits of small 
bodies in a time span comparable to their lifetimes. Two types of temporary 
stabilization were observed: short period ($\sim10^{4}$ years) when a body was in a MMR
with only one planet and long period (over $10^{5}$ years) when a body is
located in overlapping MMRs with two or three planets. 

The experiment showed that the Main Belt region can be enriched by 
cometary material in its pre-active state due to temporary resonant 
interactions between small bodies and giant planets. 

}

\maketitle

\section{Introduction}

Commensurabilities in a mean motion are very common in the Solar System. 
Resonant relations in the Main Belt have been known for nearly one and a half 
century - gaps in the distribution of asteroids were first linked to mean
motion resonances (MMRs) by Kirkwood (\cite{kw}). The influence of commensurabilities 
on the dynamics of small celestial bodies can be found in every region of the 
Solar System: from the Kuiper Belt (Nesvorny \& Roig \cite{nr1}) to the 
vicinity of the Venus orbit (Milani \& Baccili \cite{mib}).

Resonant relations of small bodies in the region among giant planets have been 
studied for the last three decades. Long time orbital stability was found close to the $L_{4}$ 
and $L_{5}$ Lagrangian points of Jupiter (Everhart \cite{eve}), in the 
narrow bands centered on 7.02 and 7.54 AU (Franklin et al. \cite{fls}) and
in the region between 24 - 27 AU (Holman \cite{hol97}). Recent papers 
describe the local (Grazier et al. \cite{dep}) or global (Robutel \& Laskar 
\cite{dmss}) dynamics in the region of the giant planets. 

In the cited works the exploration of small body dynamics was used to 
find stable regions of the Solar System where particles can survive on Gyr 
timescales on circular or quasi-circular orbits. However, the aim of this work 
is to investigate the dynamics of real particles in order to find the possible 
paths of their evolution and to learn about the role of MMRs as the mechanism 
of the orbital temporary stabilization 
process. 

\section{Method of computations}

In most of the previous papers on the subject, the results were based on the 
integrations of the equations of motion for fictitious test particles, with 
initial orbital elements chosen 'ad-hoc'. In this paper, on the contrary, 
we simulate the evolution of the orbital elements of real Centaurs and their 
clones. We chose orbits of all objects situated entirely inside the giant 
planets region: 1998 $SG_{35}$, 1999 $UG_{5}$, 2000 $EC_{98}$, 
2000 $GM_{137}$ and 2001 $BL_{41}$. All these objects are unnumbered, often 
single-oppositional which means that their elements stored in public 
databases can be approximated. That is why we decided to re-calculate their 
orbits using all available observations. Unfortunately, the data did not
allow us to acquire the elements of 2000 $GM_{137}$ with a reasonable precision, 
so we omitted this object in our computations. 

To compute the orbital elements of these real Centaurs at the initial epoch
of our simulations, we took the observations from the files available at Minor 
Planet Center in Cambridge, USA.
The sets of apparitions covered a very short time interval of about a month to a
month and a half for the 1998 $SG_{35}$, 2000 $EC_{98}$ and 2001 $BL_{41}$, 
and over 4 months for the 1999 $UG_{5}$. 
Observations were selected according to the mathematically 
objective criteria elaborated by Bielicki and Sitarski (\cite{bs1}),
assuming a normal distribution of their random errors. Only 1 observation of 
the 1998 $SG_{35}$  and 4 observations of the 1999 $UG_{5}$ were rejected, 
the rest were used in the orbit improvement process. Calculations were 
executed using Sitarski's orbital programme package (Sitarski \cite{prof2, 
prof3, prof4}). Table 1 gives the elements acquired for the bodies.

\begin{table*}
\centering
\small{
\caption{Orbital elements and their mean errors of the four bodies derived from the observational data on the epoch 1998 07 06 ET}
\begin{tabular}{rccc}
\hline
\hline
Object         &  $q$$\pm$$dq$ [AU] & $e$$\pm$$de$  & $i$$\pm$$di$ [deg] \\
RMS [arcsec] (number of obs.)   &  $\omega\pm$$d$$\omega$ [deg] & $\Omega\pm$$d$$\Omega$ [deg] & $T\pm$$dT$ [ET] \\
\hline
1998 $SG_{35}$ & 5.84069270$\pm$0.00028536 & 0.30587190$\pm$0.00001753 & 15.62341$\pm$0.00011 \\
0.38 (49 observations) & 337.90443$\pm$0.00385 & 172.49261$\pm$0.00008 & 2008 02 10.31886 $\pm$ 0.9552 \\
\hline
1999 $UG_{5}$ & 7.46349029$\pm$0.00027588 & 0.41436520$\pm$0.00006159 & 5.58734$\pm$0.00006 \\
0.61 (165 observations) & 288.36650$\pm$0.01335 & 87.02353$\pm$0.00058 & 1998 11 29.32162 $\pm$ 0.21010 \\
\hline
2000 $EC_{98}$ & 5.79188701$\pm$0.00393862 & 0.45879307$\pm$0.00049997 & 4.35728$\pm$0.00068 \\
0.48 (38 observations) & 163.21158$\pm$0.19737 & 172.62330$\pm$0.00091 & 2015 05 05.81412 $\pm$ 15.12784 \\
\hline
2001 $BL_{41}$ & 6.84577522$\pm$0.21428445 & 0.30172882$\pm$0.02988279 & 12.50754$\pm$0.16842 \\
0.49 (29 observations) &  131.62288$\pm$6.14676 & 280.58723$\pm$0.18671 & 1998 01 20.61111 $\pm$ 68.56299 \\
\hline
\end{tabular}
}
\end{table*}

\subsection{Particle cloning}

These small bodies move in the giant planets region, with the perihelion
distance close to Jupiter orbit and with the aphelion between the orbits of 
Saturn and Uranus. Orbits of the particles are chaotic due to strong planetary 
perturbations, their Lyapunov time is short and lasts about 1000 years. 

As an example, let us take two orbits of bodies moving in the gravitational 
field of the 4 planets - their 5 orbital elements are identical and they only 
differ in the starting semi-major axes by 1.5 km ($10^{-8}$ AU). Fig. 1A 
shows the eccentricity variations as a function of semi-major axis for two 
orbits - the nominal (crosses) and the changed one (circles) in a time span 
of only 11,000 years. Fig. 1B shows the logarithm of the divergence of semi-major 
axes between these orbits. The divergence is close to $10^{-8}$ AU in the 
first 3,000 years of the evolution but after this time it changes rapidly and 
both orbits start to evolve differently (see the huge leap of values of the 
logarithm on Fig. 1B). 
This means that even very small changes of one of the elements make the 
particle's dynamics quite different after a small time span. 
To avoid this problem, an ensemble of possible orbital elements is often used 
in such cases - the object's dynamical evolution is described in terms of 
probability. 

\begin{figure}[h]
\centering
\includegraphics[height=6cm]{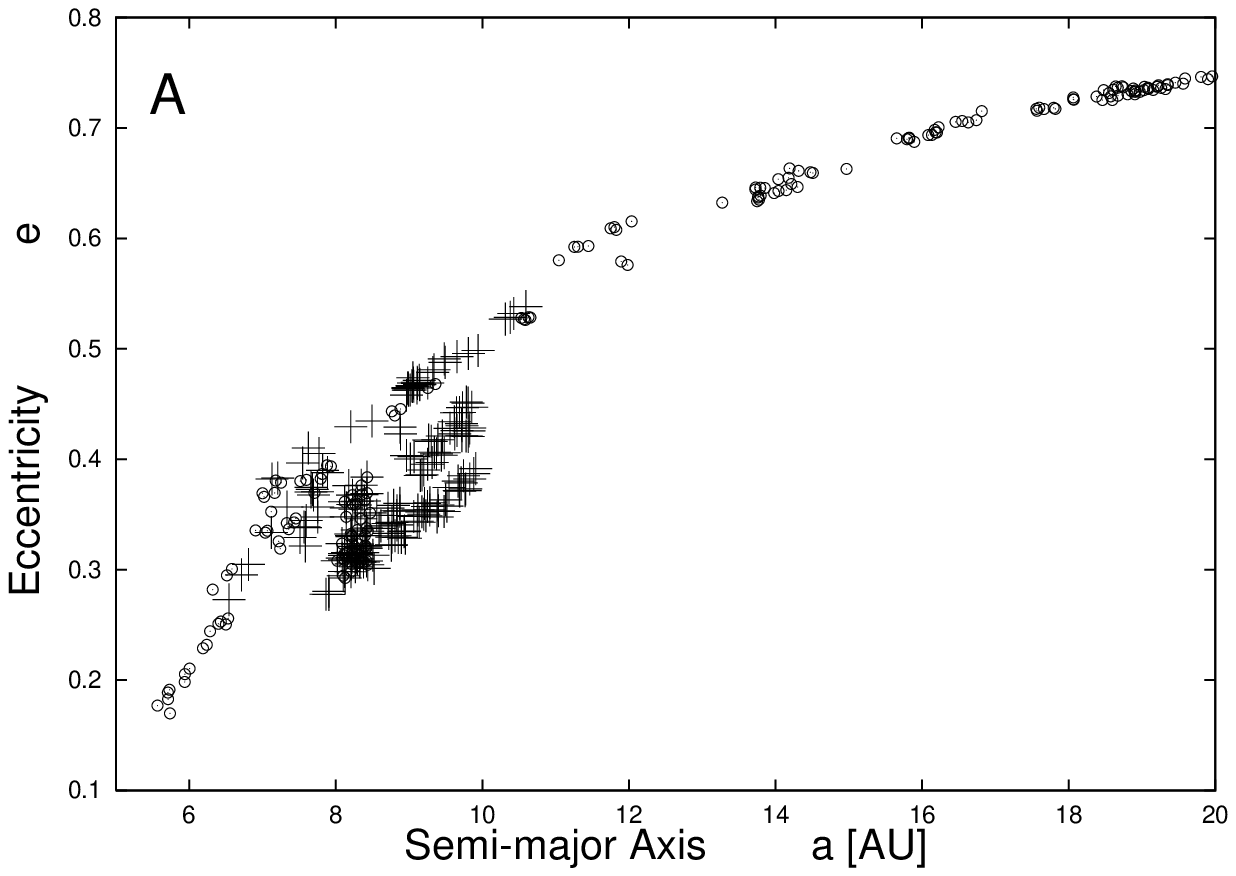}
\includegraphics[height=6cm]{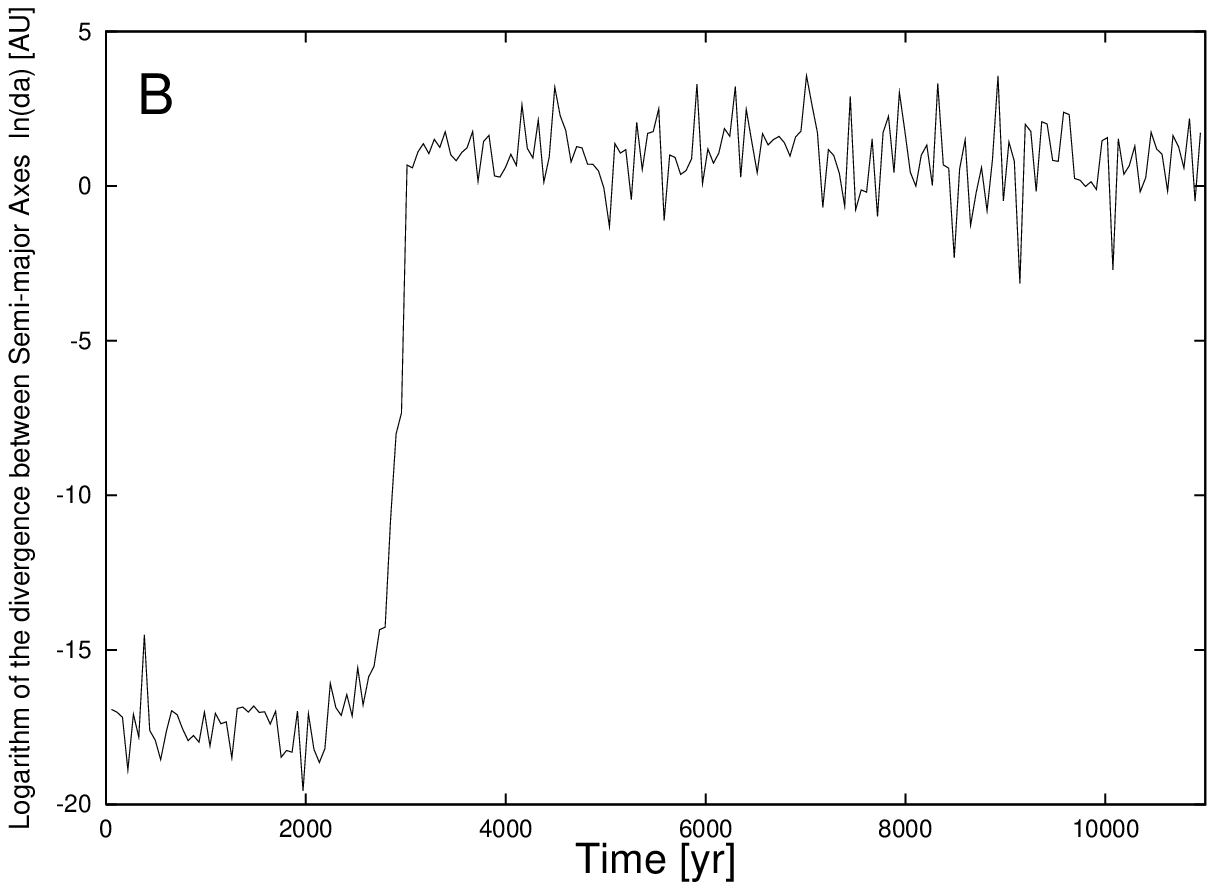}
\caption{{\bf A.} The phase space of the eccentricity variations in a
function of semi-major axis for two nearly identical orbits: the reference (crosses) and 
the changed one (circles). They differ only in the starting semi-major axes by a 
factor of $10^{-8}$. The integration time is $1.1\cdot10^{4}$ years. {\bf B.}
Time variations of the logarithm of the divergence in semi-major axes between 
orbits presented in the A. The leap near T=3000 is caused by two close 
encounters with Saturn of both objects.}
\label{fig1}
\end{figure}


Usually sets of elements are acquired by varying initial orbital elements 
within a reasonably small range, in most cases in the range of their mean 
errors. This procedure forms different orbital element sets but they cannot 
be treated as element sets of one celestial body. In fact they represent 
similar orbits of different bodies since these orbits do not fit
the observations well. The sets of elements, used in the modelling and 
described in this paper, were generated in a different way. We used the 
method (found by Sitarski \cite{sit}) for creation of orbits which represent the 
observations well. This method makes it possible to produce any number of 
orbital element sets and all of them fit the observations. Hence, these 
orbits can be treated as possible orbits of one celestial body. 

The dynamics of small bodies was modelled with the use of 404 orbital element 
sets created by the method mentioned above: for every real body 100 additional
orbits were created. 
Table 2 shows the boundary values of all six elements for each real 
object. 

\begin{table*}
\begin{center}
\small{
\caption{Dispersion of objects' orbital elements - boundary values chosen from the set of 100 additional orbits, separately for each single element}
\begin{tabular}{ccccccc}
\hline
\hline
$Object $ & $T$   &  $q$     &     $e$    &   $\omega$   &   $\Omega$   &   $i$  \\
\hline
 1998 $SG_{35}$ & 2008 02 10.55929 & 5.84125774 & 0.30592525 & 337.91370 & 172.49281 & 15.62380 \\                  
                & 2008 02 10.07866 & 5.83989244 & 0.30583258 & 337.89299 & 172.49240 & 15.62309 \\ 
\hline
 1999 $UG_{5}$  & 1998 11 29.85783 & 7.46422563 & 0.41450678 & 288.40122 & 87.02481 &  5.58749 \\ 
                & 1998 11 28.91046 & 7.46296851 & 0.41424401 & 288.34168 & 87.02197 &  5.58720 \\
\hline
 2000 $EC_{98}$ & 2015 06 12.06479 & 5.80080946 & 0.46019067 & 163.70350 & 172.62571 & 4.35859 \\ 
                & 2015 03 24.73689 & 5.78133382 & 0.45761563 & 162.65756 & 172.62160 & 4.35549 \\
\hline
 2001 $BL_{41}$ & 1998 11 18.53404 & 7.34292453 & 0.40165728 & 152.56526 & 281.04189 & 12.98724 \\ 
                & 1997 10 26.14988 & 6.23554911 & 0.24252707 & 119.25777 & 280.07133 & 12.10321 \\
\hline
\end{tabular}
}
\end{center}
\end{table*}

\section{Numerical Model}

Dynamical evolution was investigated by integrating a 6-body problem, with 
the central mass, 4 giant planets from Jupiter to Neptune, and a massless 
particle. We analyzed 404 sets of initial conditions, each of them consisting 
of positions and velocities for the planets (the same values in all cases) 
and for the massless particle. The orbital elements for the planets were taken 
from the Bretagnon's planetary theory. For the massless particle, we used the 
4 orbits of real Centaurs and 400 orbits cloned using Sitarski's method (100 
clones for each real body). The initial positions and velocities were reduced 
to the barycentre of the inner Solar System. The sets were integrated 
separately 200,000 years forwards using the 15-th order RADAU integrator 
(Everhart \cite{rad2}, \cite{rad}) with an error tolerance set to $10^{-14}$. 
This integrator adjusts the step size to maintain the accuracy for all objects 
taking part in the process. 

Locations of mean motion resonances were simply calculated with the use of 
Kepler's third law. The critical angle of the resonance was defined as follows: 

\centerline{$\sigma=k_{1}\cdot\lambda_{p}-k_{2}\cdot\lambda+(k_{2}-k_{1})\cdot\tilde\omega$}

\noindent
where $k_{1}, k_{2}$ are integers, $\lambda_{p}$ and $\lambda$ are the mean
longitudes of a planet and a test particle respectively, and $\tilde\omega$
denotes the longitude of the perihelion of test orbit. During the experiment 
only two-body MMRs were examined. 

We accepted a special method for a resonance description. In our notation, a 
resonance 2/3 means that a planet takes 2 revolutions around the barycentre 
in the time in which a small body takes 3.

\section{The influence of MMRs on small body dynamics}

The experiment showed that MMRs can temporarily stabilize numerous elliptic 
orbits in the Jupiter - Uranus region in periods of time comparable to their 
typical lifetimes (i.e., the time over which non-resonant particles 
can survive in that region). 

Two types of resonant behaviour were observed. The first one - when a small 
body is in a two-body MMR with only one planet. This relation usually lasts 
a short time, from several hundred up to 35,000 years. There were 120 particles 
of this resonant type. We only noticed two particles in a resonance with one 
planet for a longer time span. 

One of these particles, corresponding to one of the 1998 $SG_{35}$ clones, is 
shown in Fig. 2. After 118,000 years of chaotic evolution the particle enters 
the 2/3 resonance with Jupiter and remains in this relation to the end of the 
integration process. Fig. 2A presents variations of the semi-major axis as a 
function of time. The critical angle usually librates about $0^{\circ}$ 
when a body is located in this particular resonance (see Fig. 2B). 
Even high and growing amplitude (over $150^{\circ}$) of the librations is 
incapable of destabilizing the dynamics of the particles in a time span of 
over 80,000 years.

\begin{figure}
\centering
\includegraphics[height=6cm]{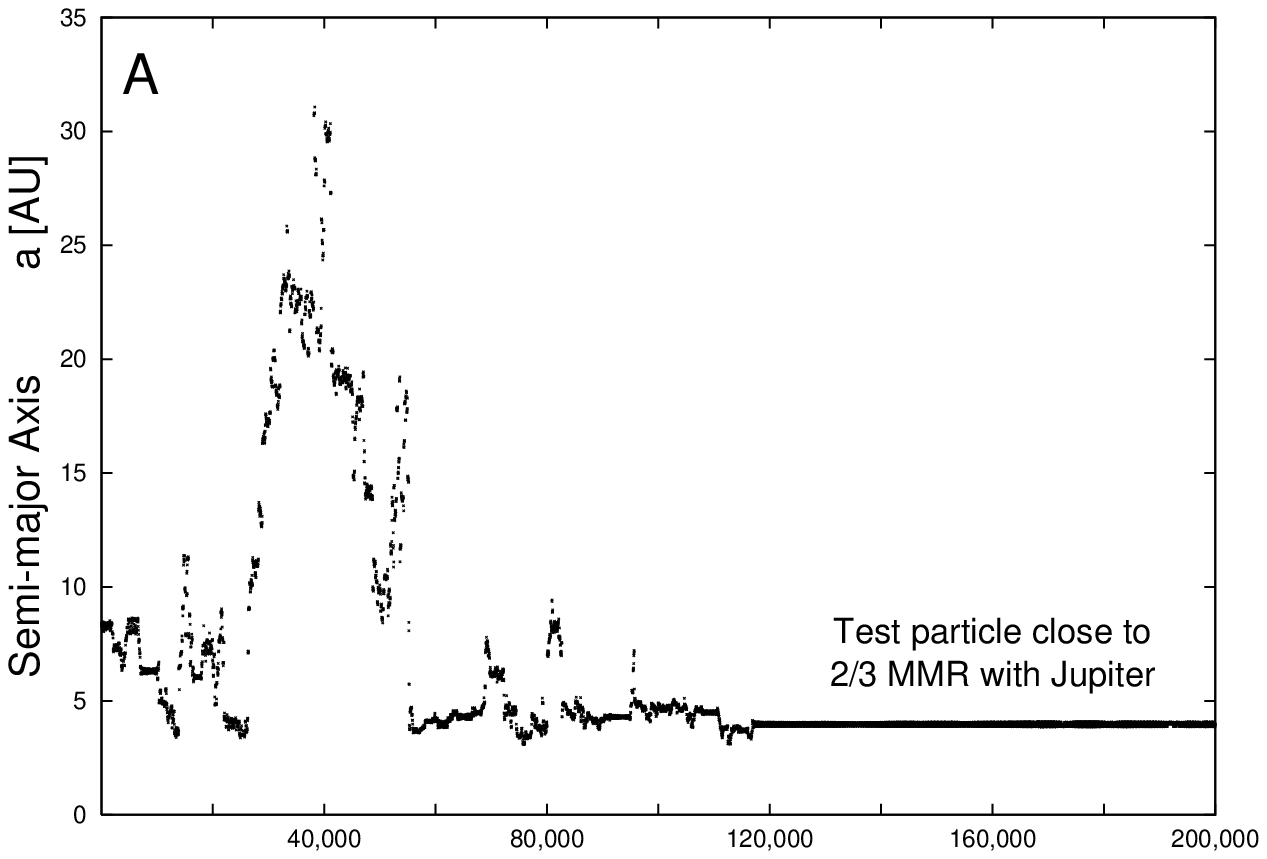}
\includegraphics[height=6cm]{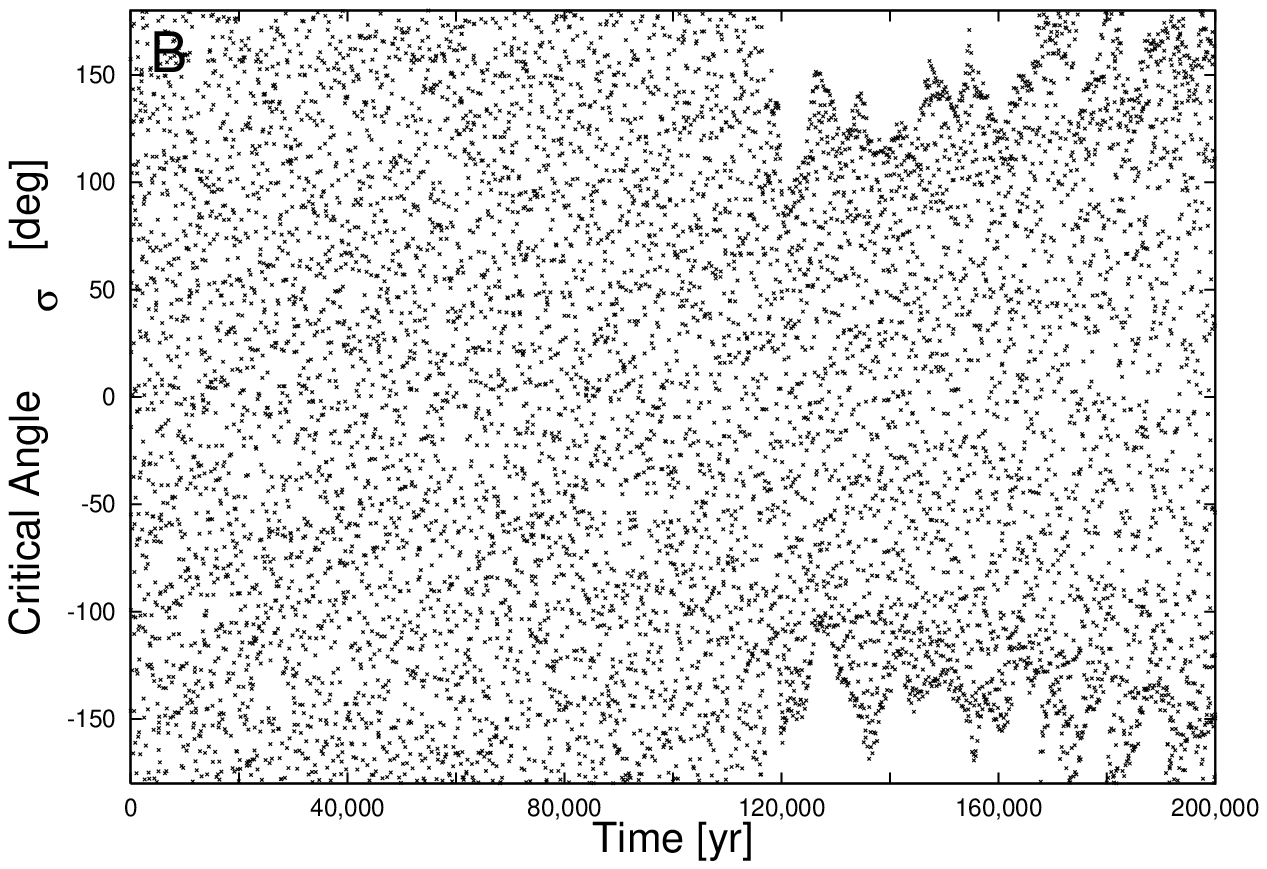}
\includegraphics[height=6cm]{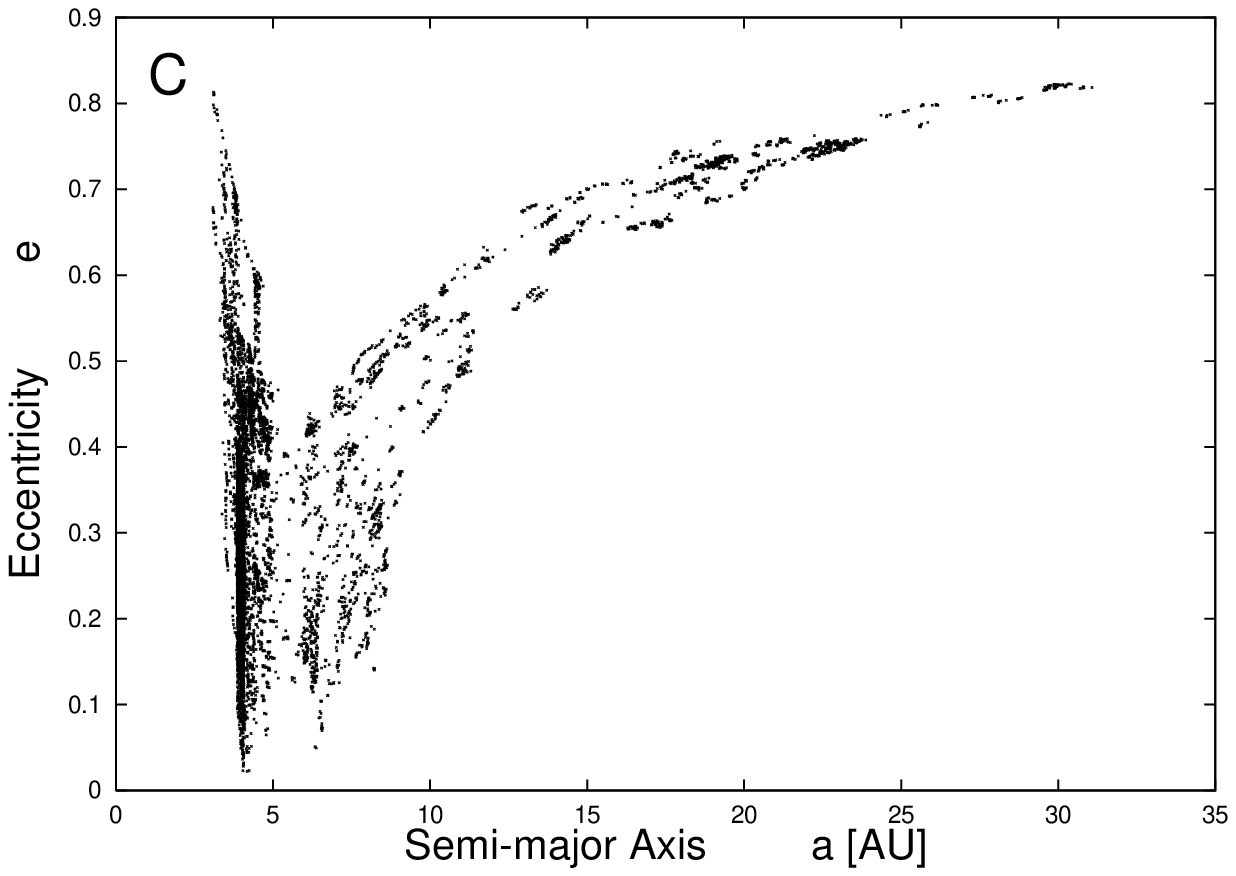}
\caption{1998 $SG_{35}$ - test particle no 35. {\bf A.} Variations of the 
semi-major axis as a function of time. The orbit is locked in a 2/3 MMR with 
Jupiter after 118,000 years of dynamical evolution. {\bf B.} Variations 
of the critical angle as a function of time. When the particle is in resonance, 
the critical angle starts to librate about $0^{\circ}$ with a large amplitude 
(dotted-free area indicates values of angles which are unaccessible for the 
librating body). The amplitude increases in time and the libration is hardly 
visible 
at the end of the integration (dotted-free area almost disappears). The 
particle will probably leave the resonance in several thousand years. When 
the body remains outside the resonance, values of the critical angle vary at 
random. {\bf C.} The phase space of the eccentricity variations in the
function of the semi-major axis. When a body is located in a resonance, the 
eccentricity can reach high values but the particle does not leave the MMR.
The body seems to "stick" to the resonance.} 
\label{fig2}
\end{figure}
    
Similarly to this Hilda type orbit, nearly all other orbits remaining in 
resonance with only one planet behave like the Toro class orbits (see 
Fig. 3). This class defined by Milani (\cite{mi1}) consists of Earth-crossing 
objects which avoid close approaches by entering a MMR. Most particles
remaining in the resonant relation with Jupiter or Saturn cross their orbits 
in ecliptic projections but cannot approach them. The 
time of such resonant relations is smaller than for Toro like orbits (with 
some exceptions) but this can be caused by stronger perturbations acting on 
a body in the giant planet region. 

\begin{figure}
\centering
\includegraphics[height=6cm]{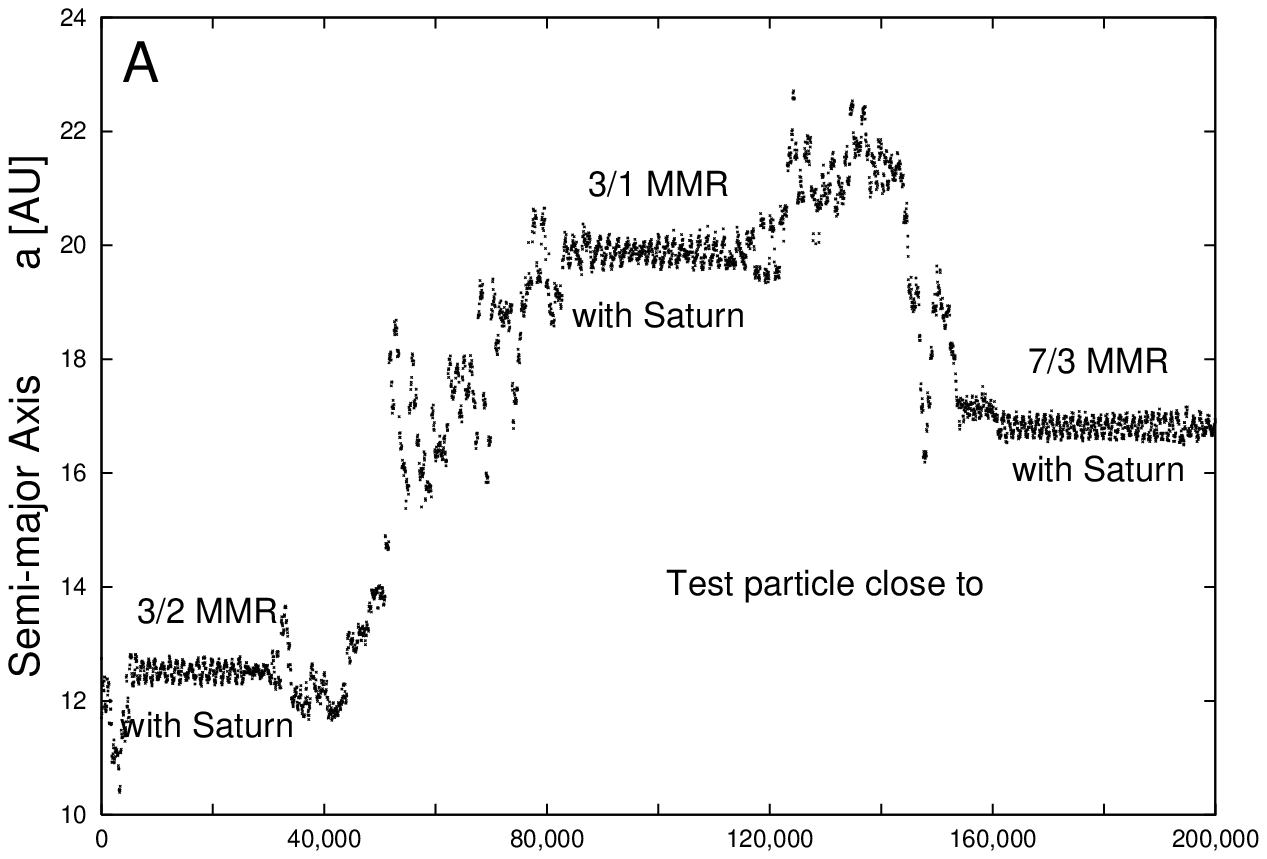}
\includegraphics[height=6cm]{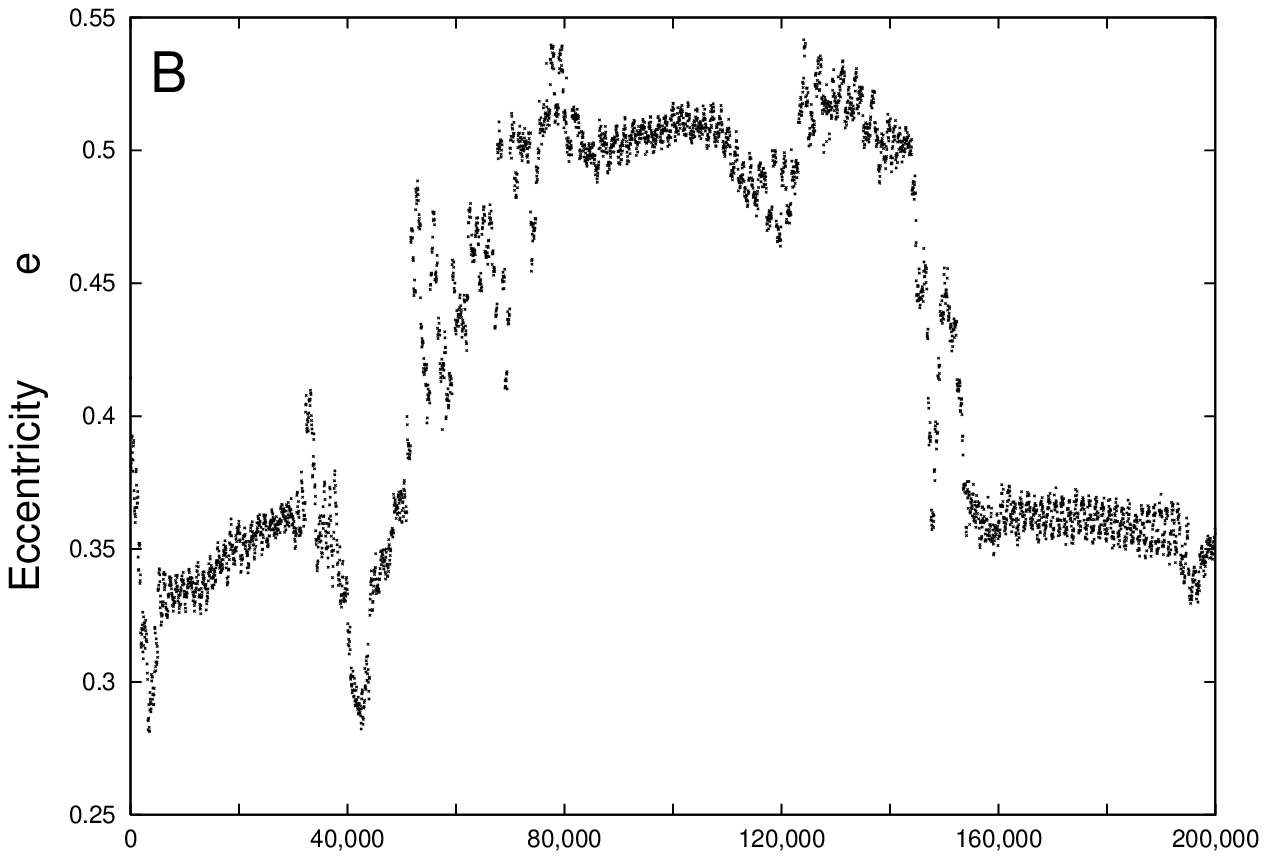}
\includegraphics[height=6cm]{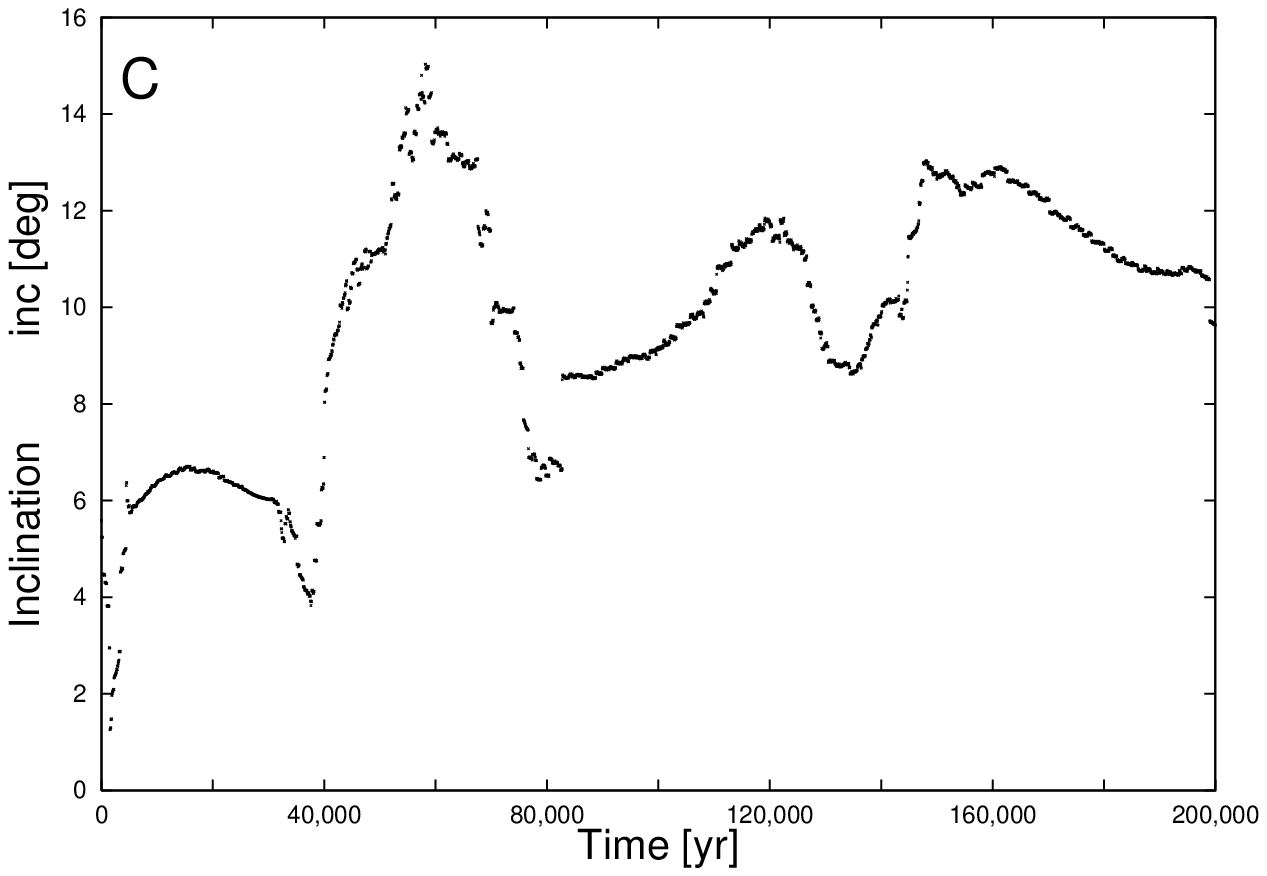}
\caption{1999 $UG_{5}$ - test particle no 1. {\bf A.} During its dynamical
evolution, the particle enters different MMRs. When in resonance, the variations 
of eccentricity ({\bf B.}) and inclination ({\bf C.}) values evolve slowly 
(but not periodically)  
in small boundaries due to the lack of close approaches to planets. Most of
these orbits are Toro class, which means that they cross planetary orbits (in
ecliptic projections) but avoid close approaches to these massive bodies.} 
\label{fig3}
\end{figure}

The second type of resonant behaviour is characterized by much longer 
resonant interactions up to 160,000 years in our sample. Test particles 
are in two-body MMRs with 2 or 3 planets (see Fig. 4). These overlapping resonant 
relations can stabilize orbits far more effectively - for over $10^{5}$ years. 
Fig. 4 presents the evolution of the MMRs critical angles for one of
these orbits. When two MMRs overlap each other, the behaviour of the
critical angle of the dominant resonance starts to vary (see Fig. 4C) - the 
critical angle begins to circulate with an overlapped libration of amplitude 
almost constant.
Irregularities of the rotation and of the amplitude of libration 
do not mean that the body will leave the MMRs in the near future
but the possibility that a small body will approach a planet increases with 
time. 
Only 9 to 15 \% of orbits were stabilized on time spans longer than 35,000 
years (according to the object). 

Fig. 4 shows that overlapping resonances can also protect the body
from close approaches to planets. The disturbing resonance does not 
necessarily destabilize the orbit of the body if the influence is weak. 

\begin{figure}
\centering
\includegraphics[height=5cm]{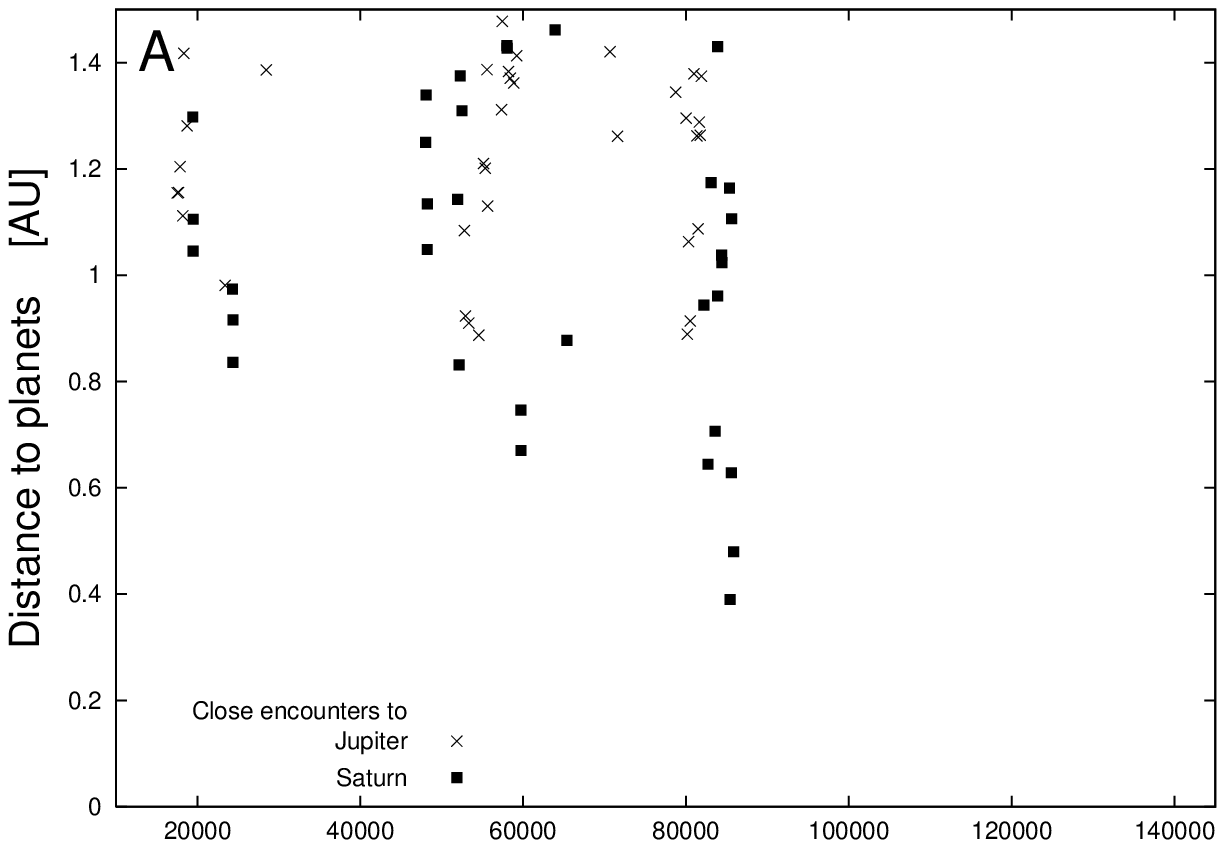}
\includegraphics[height=5cm]{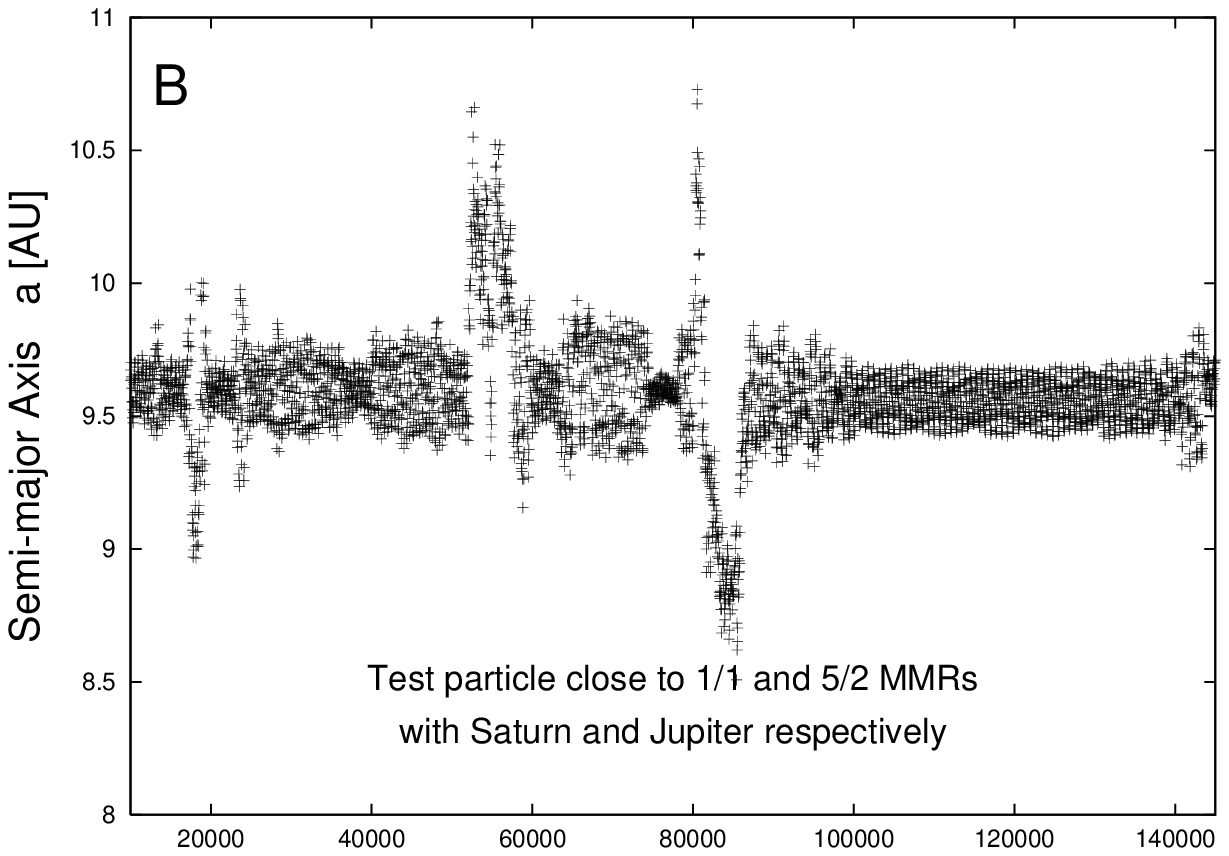}
\includegraphics[height=5cm]{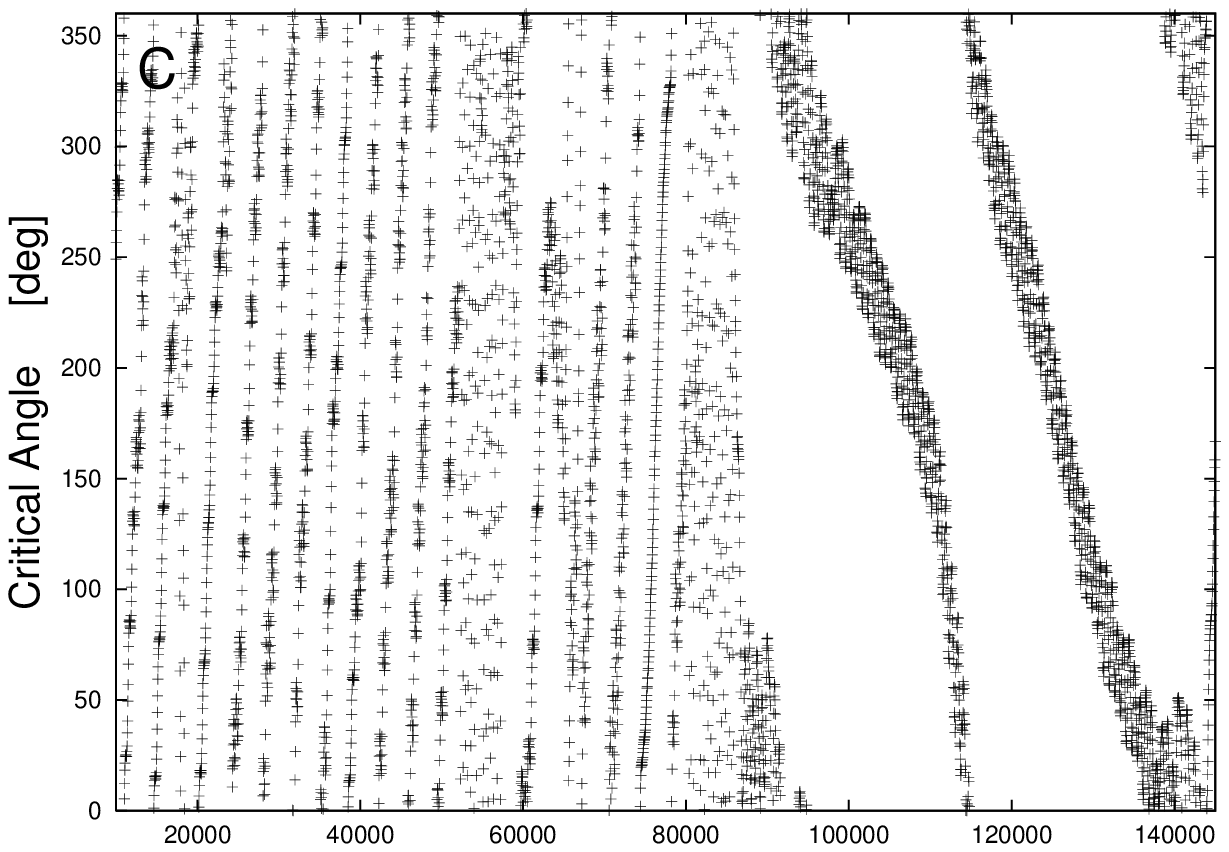}
\includegraphics[height=5cm]{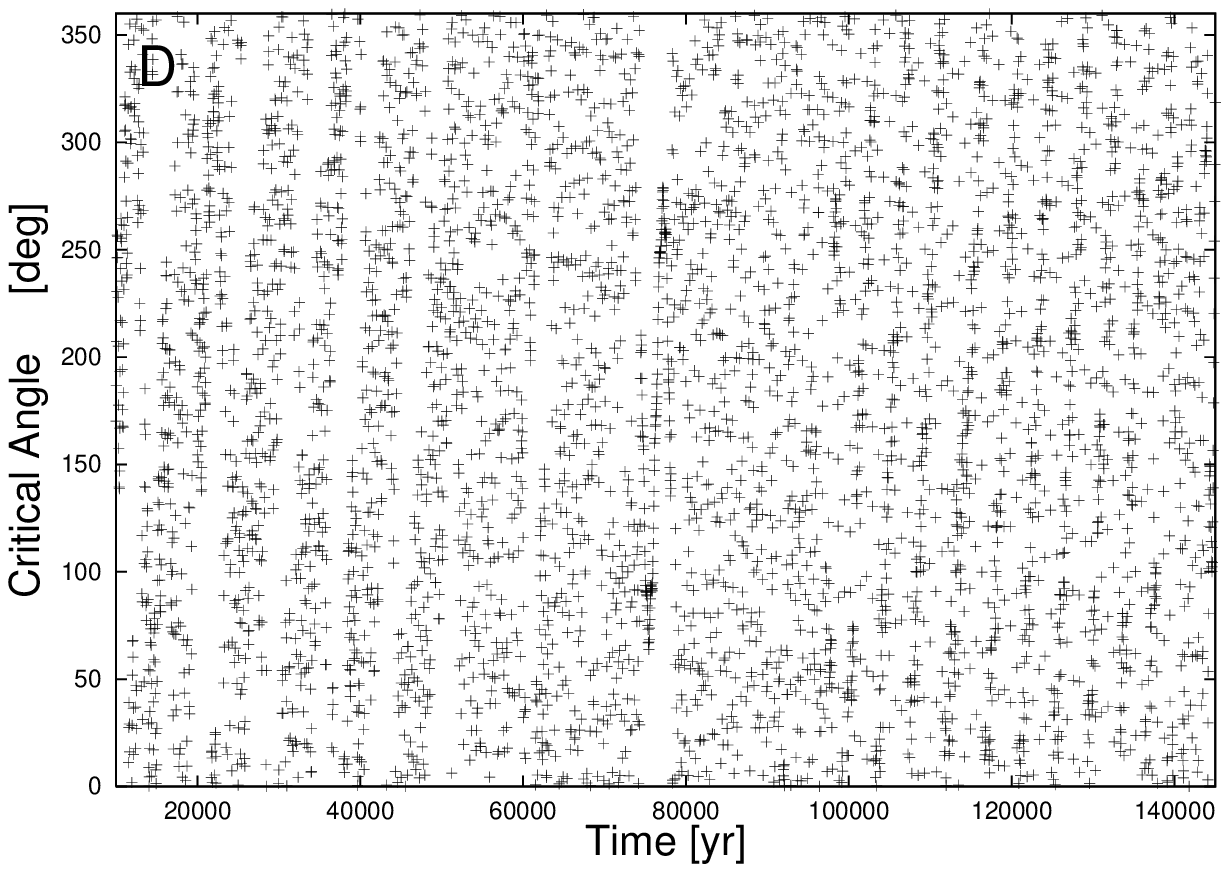}
\caption{2001 $BL_{41}$ - test particle no 11. {\bf A.} The distance between 
the body and planets as a function of time. {\bf B.} Variations of the 
semi-major axis as a function of time. The orbit is located in overlapping 
MMRs due to near resonant relations between Jupiter and Saturn (so called 
Great Inequality). Time variations of the MMR critical angles of Saturn ({\bf 
C}) and Jupiter ({\bf D}) are also presented. After 100,000 years of
dynamical evolution, when the body remains in 1/1 MMR with Saturn and in  
5/2 disturbing MMR with Jupiter, it has no close approaches to both planets.}
\label{fig4}
\end{figure}

\section{Migration to the Main Belt}

About 30 - 40 \% of test bodies migrate to the Main Belt on non-cometary 
(asteroidal) orbits (see Hilda type orbit in Fig. 2). Most of them leave this 
region in a dynamically short time, usually in less than 250 orbital 
periods without entering a MMR, but about 30 \% of them will have resonant 
interactions with Jupiter and/or Saturn. 
When they enter a MMR, their lifetime in the Main Belt grows 
rapidly - up to 20,000 orbital revolutions in our calculations. 

The modelling shows that proto-cometary objects in their pre-active
state can enrich the Main Belt population. A particularly interesting 
case is that of the short-lived resonant groups of asteroids, like the 
Griquas. This is a group of resonant asteroids dynamically differentiated from 
the other resonant groups, and that cannot be primordial resonant bodies due 
to their short lifetimes. These vary from $10^{4}$ to $10^{7}$ years (Roig et 
al. \cite{ro1}) which may indicate that a certain per cent of these asteroids 
came from another regions. Have they come from the outer Solar System ? 
Unfortunately, our integrations do not give an answer to this question. The 
calculations indicated only 1 particle with $3.1 < a < 3.27$ and $e > 0.35$ 
close to a 1/2 MMR with Jupiter with the orbital lifetime longer than $10^{4}$ 
years.

We decided to integrate backwards the equations of motion of 147 
real asteroids moving close to 1/2 MMR with Jupiter to find out if 
these objects are able to come from the region outside the orbit of Jupiter. 
The results shows that 1/2 MMR members can be enriched by Centaurs coming from 
two different regions but the backward simulations cannot be treated as the 
evidence. 

\section{Islands of temporary stability}

The numerical experiment allowed one to find temporary "islands of stability" 
in the chaotic region of the giant planets - bands where particles 
are trapped for the time estimated for $10^{4}-10^{5}$ years. 
These "islands" are observed near 6, 6.8, 8.2, 9.5, 10, 10.5, 11, 11.6, 12.1, 
13, 15.1 and 17.3 AU in the giant planets region, and also close to 2.8 - 3.1 
and 3.4 - 4.2 AU in the Main Belt. 

\begin{figure*}
\centering
\includegraphics[angle=180,width=15cm]{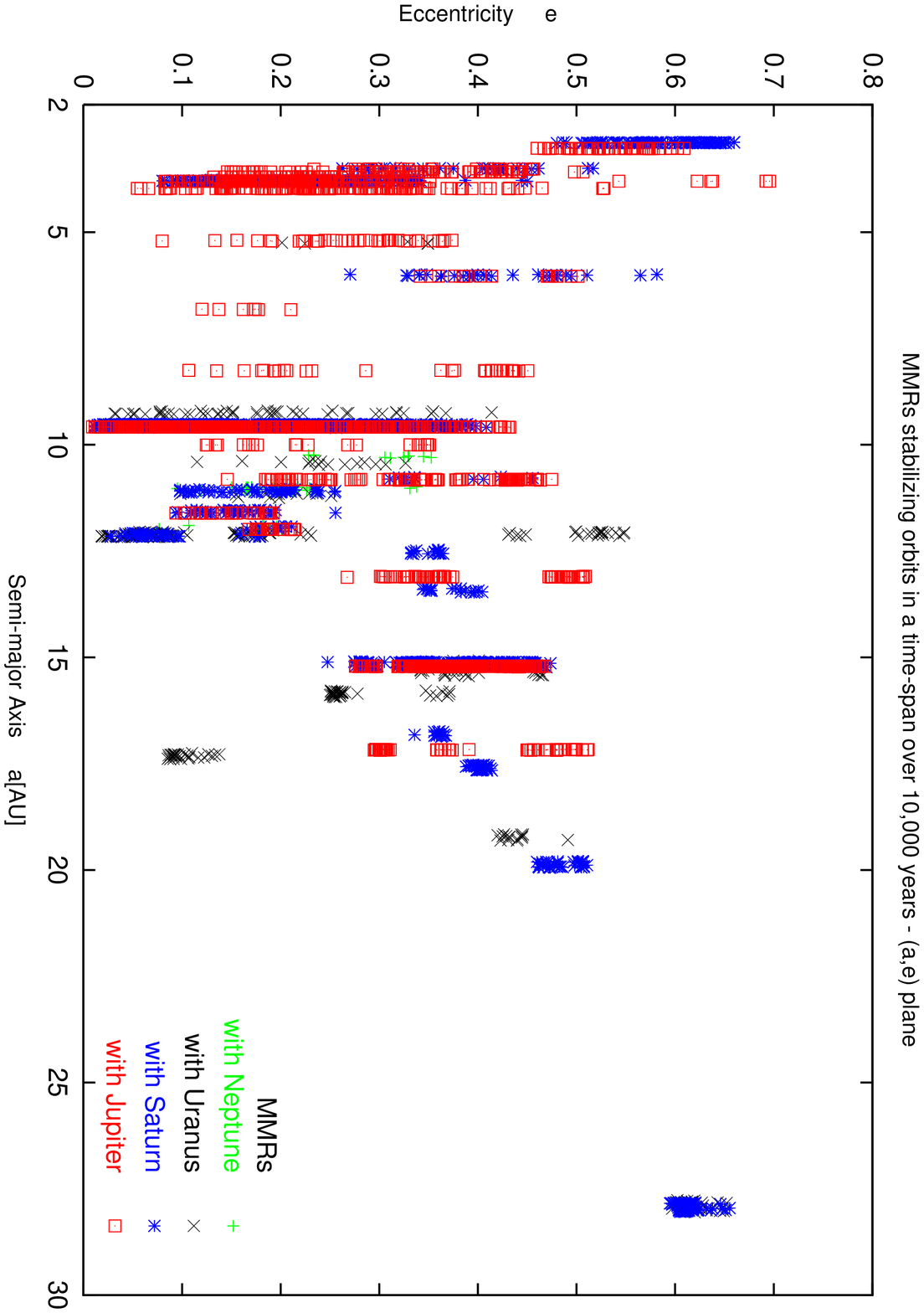}
\caption{"Islands of temporal stability" among the giant planets in the 
(a,e) plane. High eccentricities do not prevent MMRs from stabilizing 
orbits in a time span longer than $10^{4}$ years due to stickness of 
resonances. The region outside Uranus seems to be empty but this can be 
caused by the ensemble selection.}
\label{fig5}
\end{figure*}

Fig. 5 shows the bands of stability in the ($a$,$e$) plane. All orbits 
located in these zones were temporarily stabilized by MMRs in the time span 
of over $10^{4}$ years. This period of time is comparable with the 
mean lifetime of the non-resonant particles in the region among giant 
planets estimated by Franklin et al. (\cite {fls}).
Most of the resonances are observed in the Main Belt and outside Saturn's 
orbit. The region between Jupiter and Saturn is weakly inhabited. This 
can be caused by strong gravitational perturbations from these two planets. 
The other hardly inhabited region (in fact nearly empty) is located outside 
the orbit of Uranus. This can be due to the ensemble selection. 

The most effective MMRs stabilizing dynamics of particles are gathered in 
Table 3. 

\vskip 0.3cm
\begin{table}
\small{
\caption{Frequently observed MMRs and their stabilization time}
\begin{center}
\begin{tabular}{rcl}
\hline
\hline
Object         &  MMR(s)  & time of orbital \\ 
               &          &  stabilization \\
\hline
1998 $SG_{35}$ & 2/3 (J)             & 80 kyrs \\
               & 2/1 (S) and 5/1 (J) & 50 kyrs \\
               & 1/1 (S) and 5/2 (J) & 50 kyrs \\
               & 6/7 (U), 1/3 (U), 4/9 (N) &  20 kyrs \\  
\hline
1999 $UG_{5}$  & 2/1 (S) and 5/1 (J) & 60 kyrs \\
               & 1/4 (S) and 5/8 (J) & 35 kyrs \\
               & 5/4 (S) and 2/9 (N) & 25 kyrs \\
               & 1/1 (S) and 5/2 (J) & 20 kyrs \\
               & 4/7 (J)             & 10 kyrs \\  
\hline
2000 $EC_{98}$ & 1/4 (S) and 5/8 (J) & 120 kyrs \\
               & 5/2 (J), 1/1 (S) and 1/3 (U) & 110 kyrs \\
               & 4/9 (J)             &  65 kyrs \\
               & 1/1 (S) and 5/2 (J) & 30 kyrs \\
\hline
2001 $BL_{41}$ & 2/1 (S) and 5/1 (J) &  150 kyrs \\
               & 2/1 (S), 5/1 (J) and 5/7 (U) & 90 kyrs \\
               & 1/1 (S) and 5/2 (J) & 70 kyrs \\
\hline
\end{tabular}
\end{center}
}
\end{table}

\section{Discussion and Conclusions}

The region among the giant planets is one of the most chaotic in the Solar
System. Strong perturbations cause small bodies to evolve in the region 
among giant planets over a dynamically short time equal to the 
$10^{3} - 10^{4}$ years. Our results show that this time can be much longer 
if the body is temporarily trapped by a MMR with a planet. 

Temporary stable orbits of the Centaurs seem to be a good example of the 
"resonance stickness" (Dvorak et al. \cite{dvo}, Tsiganis et al. \cite{tsg}). 
Gravitational perturbations can easily push a body into a MMR even if the 
eccentricity of the body is high. The evolution of the semi-major axis, eccentricity 
and inclination is chaotic in the borders of the resonance. The perturbations 
can push the body outside the resonance in timescale of $10^{4} - 10^{5}$ 
years in the same way as they pushed the object into this MMR. The number of small bodies 
which enter a resonance on highly eccentric orbits during a short time 
of dynamical evolution (200,000 years) indicates that a resonance sticking is 
a common mechanism affecting particle lifetimes in chaotic zones. 

To ensure that the phenomenon we encountered is in fact the resonance stickness, 
we investigated the autocorrelation function of the time series of the osculating 
elements of one of the resonant orbits. If the body is "stuck" to a resonance 
we observed quasi-periodic variations of orbital elements on a 
timescale much longer than the Lyapunov time of this orbit (Tsiganis et 
al. \cite{tsg}). 

Fig. 6 presents the autocorrelation function of the semi-major axis as a
function of the time-lag k for one of the test particles being in 1/1 and 5/2
MMRs with Saturn and Jupiter, respectively. Time series consisted of 1302 
points (which corresponds to 40,000 years) starting at t=100,000 years. 
The series is strongly suppressed in time but this is due to the presence of 
gravitational perturbations from the giant planets in our 6 
body problem. If we had a 3-body problem the suppression would be absent. 
Although the suppression is high, quasi-periodic variations
can be observed. This seems to confirm the thesis of resonance stickness. 

\begin{figure}[h]
\centering
\includegraphics[width=9cm]{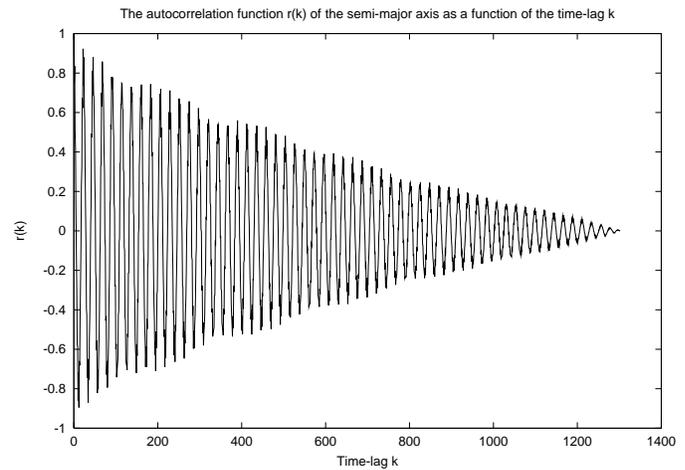}
\caption{2001 $BL_{41}$ - test particle no 11. The autocorrelation function 
r(k) of the semi-major axis as a function of the time-lag k. Time series 
consists of 1302 points which correspond to 40,000 years of time.}
\label{fig6}
\end{figure}

The most unintuitive result is that overlapping resonances can stabilize
small body orbits on a much longer timescale than single resonances. The
overlapping resonances are usually responsible for chaotic evolution, but
our results show that they can stabilize dynamics rather than destabilize it. 
When we observe the particle motion in overlapping resonance with 
planets, we can see that its evolution is chaotic but only inside
the borders of the MMRs. The resonances protect the body from close
approaches to planets. Giant planets are close to resonant relations 
(Jupiter - Saturn - Uranus and Saturn - Uranus - Neptune, Sussmann and Wisdom 
\cite{sc2}, \cite{sc3}) which means that a small body can be locked in a 
MMR with 1 planet and close to MMRs with the other 1 or 2 massive bodies. When 
gravitational perturbations change the orbit of the particle, the body can be 
easily "pumped up" into another MMR. Phase space of MMRs is thicker when the 
eccentricity increases and hence bodies can remain in overlapping MMRs. 
The overlapping resonances protect the body from close approaches to 
planets (which are the main mechanism of chaos in the region)  and according
to our calculations bodies are unable to change the portion of the phase 
space where they evolve, on a timescale of over $10^{5}$ years. 

Resonances between the test particle and the 4 planets have not been 
observed. This is in good agreement with Sussman's and Wisdom's works cited 
above. 

The results presented in this work indicate that Centaurs may also enrich 
the population of asteroids in the Main Belt. MMRs can stabilize Centaur 
orbits on the timescale of $10^{4}$ - $10^{5}$ years. This is a 
very short period of time in comparison with the lifetime of some 
asteroidal families in this region but shows that the Main Belt can include 
non-processed inactive cometary nuclei, not only extinct material. It is hard to say 
if such nuclei can be called primordial but our results show that the material 
can reach the asteroid belt on non-cometary orbits, without approaching the Sun. 
The particles did not enrich specific groups of asteroids in the Main Belt.  
The only group which seems to be overpopulated are Trojans. 

We tried to consider if Centaurs can be a parental population of groups 
of asteroids with the lifetime comparable to $10^{4} - 10^{5}$ years. Griquas 
are an example of such a group. They do not originate in the 
break-up event, a lot of the group members have lifetimes comparable 
to the lifetimes of Centaurs in the Main Belt. But, unfortunately, our 
modelling did not produce a group of objects with $3.1 < a < 3.27$ and 
$e > 0.35$. Although, when we integrated backwards all known orbits of the
Griqua group members, we found that some of them are able to come from 
the regions among giant planets. Further investigations are needed to explain 
the origin of Griqua type bodies. 

The earlier works of Everhart (\cite{eve}) and Franklin et al. (\cite{fls}) 
indicate that there are two narrow bands at 7.02 and 7.54 AU where particles 
can survive longer than $10^{6}$ years. In 1997 Holman (\cite{hol97}) pointed
to a region between 24 and 27 AU where low inclined ($i\leq1^{\circ}$) and low 
eccentric ($e\leq0.01$) objects can survive 1 Gyr. Our results do not
indicate any stable orbits close to those bands. This means that 
the resonant dynamics is not the main mechanism responsible for stabilizing 
orbits in these areas. Slow evolution of low inclined, quasi circular orbits 
close to bands at 7.02 and 7.54 AU and in the Holman's region is caused by 
tiny variations of the gravitational force acting on bodies. 

Centaurs are perceived as a intermediate stage between Kuiper Belt objects and
Jupiter Family comets (Levison \& Duncan \cite{levi}). During the
integration we noticed only 10 \% of orbits with $q < 1.5$ AU and $Q > 5.0$ AU.
Most of them were orbits of 1998 $SG_{35}$ and 2001 $BL_{41}$ objects (79
and 13 orbits respectively). 

It is hard to compare our results with those presented in Grazier's 
and Robutel's papers due to different timescales. Their work focused on 
research of long time stability of Solar System regions while we studied the 
paths of a small ensemble of orbits. We believe that the results 
of our experiment and Robutel's survey would be similar for the same 
timescales of integrations. 

Further studies are required to broaden our knowledge on orbital stabilization 
by overlapping and many-body MMRs in the chaotic region of the Solar System. 

\begin{acknowledgements}
We thank G. Sitarski for help and comments. Many thanks to the referee 
F. Roig for his valuable discussions and helpful reviews. The computations 
were accomplished in the Pozna{\'n} Supercomputing and Network Centre and in 
the Interdisciplinary Centre of Mathematical Modelling of the Warsaw 
University.
\end{acknowledgements}


\end{document}